\documentclass[aps,pra,twocolumn,amsmath,amssymb,nofootinbib,superscriptaddress]{revtex4}
\usepackage[english]{babel}
\usepackage{latexsym}
\usepackage{graphics}
\usepackage{epsfig}
\usepackage{color}
\usepackage{bm}
\usepackage{amsmath}
\usepackage{amssymb}

\usepackage[normalem]{ulem}
\usepackage{color}

\newcommand{\1}{\uparrow}
\newcommand{\2}{\downarrow}

\begin{document}

\title{Ultradilute low-dimensional liquids}
\author{D.~S.~Petrov}
\affiliation{LPTMS, CNRS, Univ. Paris Sud, Universit\'e Paris-Saclay, 91405 Orsay, France}
\author{G.~E.~Astrakharchik}
\affiliation{Departament de F\'isica, Campus Nord B4-B5,
Universitat Polit\`ecnica de Catalunya, E-08034 Barcelona, Spain}


\begin{abstract}

We calculate the energy of one- and two-dimensional weakly interacting Bose-Bose mixtures analytically in the Bogoliubov approximation and by using the diffusion Monte Carlo technique. We show that in the case of attractive inter- and repulsive intraspecies interactions the energy per particle has a minimum at a finite density corresponding to a liquid state. We derive the Gross-Pitaevskii equation to describe droplets of such liquids and solve it analytically in the one-dimensional case.

\end{abstract}


\maketitle

According to van der Waals' theory the fundamental property of a liquid to form self-bound states with free surface is due to the shape of the interaction potential which typically has
a repulsive core and a more extended attractive part. Usual liquids are dense and almost incompressible since particles prefer to be at the potential minima. A qualitatively different type of liquid, dilute one, has very recently been observed in a Bose-condensed Dy gas characterized by anisotropic dipolar interactions \cite{PfauDroplet1,PfauDroplet2} and a similar phenomenon has been predicted to occur in three-dimensional Bose-Bose mixtures with isotropic contact interactions \cite{PetrovDroplet}. In both cases the system, collapsing from the mean-field viewpoint, is stabilized by quantum many-body effects; each particle feels the attractive mean-field interaction proportional to the density $n$ compensated by the positive Lee-Huang-Yang correction $\propto n^{3/2}$ \cite{PetrovDroplet,PfauDroplet1,PfauDroplet2,SantosDroplet,Saito}. Such liquids and their finite-size droplets remain dilute and weakly interacting allowing for a well-controlled perturbative description. They also have quite peculiar features: their very existence is a direct manifestation of beyond mean-field effects, they require no trapping and their bulk density and shape are tunable by changing interactions, in the absence of external trapping they  can reach zero temperature by evaporation, etc.

In this Letter, motivated by the enhanced role of beyond-mean-field effects in low dimensions \cite{Remark2D}, we consider two- and one-dimensional Bose-Bose mixtures and show that with decreasing the dimensionality the liquid phase not only persists, but becomes more ubiquitous and remarkable. We find that in the two-dimensional case the energy per particle is proportional to $n[\ln (n/n_0)-1]$ ($n_0$ is the equilibrium density), the liquid state exists as long as the interspecies interaction is weakly attractive and the intraspecies ones are weakly repulsive. This contrasts the three-dimensional case where a critical interspecies attraction is needed to liquefy the mixture. Interestingly, we find that a three-dimensional mixture in the gas phase can become liquid if confined to the quasi-two-dimensional geometry. In the one-dimensional case the liquid phase originates from the competition of a {\it repulsive} mean-field term $\propto n$ and {\it attractive} beyond mean-field correction $\propto -n^{1/2}$. Counterintuitively, this means that a one-dimensional mixture, stable from the mean-field viewpoint, is actually unstable towards the formation of a liquid droplet. We analytically describe its shape and other properties.

Consider two equal-mass bosonic species ($\sigma =\1;\2$), with densities $n_\1$ and $n_\2$, governed by the Hamiltonian
\begin{eqnarray}\label{Ham}
H&=&\sum_{\sigma,\bf k}\frac{k^2}{2}\hat{a}_{\sigma,{\bf k}}^\dagger \hat{a}_{\sigma,{\bf k}}\nonumber\\
&&\hspace{-1.cm}+\frac{1}{2}\sum_{\sigma,\sigma',{\bf k}_1,{\bf k}_2,{\bf q}} \hat{a}_{\sigma,{\bf k}_1+{\bf q}}^\dagger \hat{a}_{\sigma',{\bf k}_2-{\bf q}}^\dagger U_{\sigma\sigma'}({\bf q}) \hat{a}_{\sigma,{\bf k}_1}\hat{a}_{\sigma',{\bf k}_2},
\end{eqnarray}
where $U_{\sigma\sigma'}$ are short-range interaction potentials and we set $m=\hbar = 1$. As usual \cite{Beliaev,Popov1971,LandauIX}, one substitutes $U_{\sigma\sigma'}$ by effective potentials, characterized by the same scattering amplitudes for relevant collision energies but more suitable for perturbative expansions.

We first discuss the two-dimensional case and take $U_{\sigma\sigma'}({\bf q})=g_{\sigma\sigma'}={\rm const}\ll 1$ for $|{\bf q}|<\kappa$ and $U_{\sigma\sigma'}({\bf q})=0$ for $|{\bf q}|>\kappa$. The coupling constants $g_{\sigma\sigma'}$ and the cutoff $\kappa$  are related to the two-dimensional scattering lengths $a_{\sigma\sigma'}>0$ by $g_{\sigma\sigma'}=4\pi/\ln(\epsilon_{\sigma\sigma'}/\kappa^2)$, where $\epsilon_{\sigma\sigma'}=4e^{-2\gamma}/a_{\sigma\sigma'}^2$ and $\gamma$ is Euler's constant. This relation ensures that at low energy, $z\ll \kappa^2$, the scattering {\it t} matrix behaves as $t_{\sigma\sigma'}(z)\approx 4\pi/\ln(-\epsilon_{\sigma\sigma'}/z)$ \cite{Popov1971,Jackiw1991} consistent with the Born series expansion $t_{\sigma\sigma'}(z)\approx g_{\sigma\sigma'}[1-g_{\sigma\sigma'}\ln(-\kappa^2/z)/4\pi+...]$. One can see that the perturbation series in terms of $|t_{\sigma\sigma'}|\ll 1$ and $|g_{\sigma\sigma'}|\ll 1$ are equivalent as long as $\kappa^2$ is larger but not exponentially larger than the typical interaction energy $z$ which is the product of the density $n$ and the {\it t} matrix (with the logarithmic accuracy one can simply use $z\sim n$). An appropriate value of $\kappa$ can always be found in the weakly interacting regime where the scattering lengths are exponentially small (repulsion) or large (attraction) compared to the mean interparticle separation.

In order to calculate the ground-state energy of the mixture up to second order terms in $g$ we do the standard Bogoliubov theory (see, for example, \cite{LandauIX}). Namely, we assume a macroscopic condensate population $\hat{a}_{\sigma,0}\approx \sqrt{n_\sigma}$, expand (\ref{Ham}) up to bilinear terms in the operators $\hat{a}^\dagger_{\sigma,{\bf k}}$, $\hat{a}_{\sigma,{\bf k}}$ for $k\neq 0$, and diagonalize the bilinear form arriving at the ground-state energy density
\begin{equation}\label{BogIntermediate}
E=\frac{1}{2}\sum_{\sigma\sigma'}g_{\sigma\sigma'}n_\sigma n_{\sigma'}+\frac{1}{2}\sum_\pm \sum_{|{\bf k}|<\kappa}[E_\pm (k)-k^2/2-c_\pm^2],
\end{equation}
where $E_\pm (k)=\sqrt{c_\pm^2 k^2+k^4/4}$ are the Bogoliubov modes with sound velocities $c_\pm$ defined by
\begin{equation}\label{cpm}
c_{\pm}^2=\frac{g_{\1\1}n_\1+g_{\2\2}n_\2\pm\sqrt{(g_{\1\1}n_\1-g_{\2\2}n_\2)^2+4g_{\1\2}^2 n_\1 n_\2}}{2}.
\end{equation}
The momentum integration in Eq.~(\ref{BogIntermediate}) gives
\begin{equation}\label{BogKappa}
E_{\rm 2D}=\frac{1}{2}\sum_{\sigma\sigma'}g_{\sigma\sigma'}n_\sigma n_{\sigma'}+\frac{1}{8\pi}\sum_{\pm}c_{\pm}^4\ln\frac{c_{\pm}^2\sqrt{e}}{\kappa^2}.
\end{equation}
Recalling that $g_{\sigma\sigma'}=4\pi/\ln(\epsilon_{\sigma\sigma'}/\kappa^2)$ one can check that to the chosen order $\partial E_{\rm 2D}/\partial \kappa^2=0$, i.e., the final result (\ref{BogKappa}) depends only on $n_\sigma$, $a_{\sigma\sigma'}$, and not on $\kappa$.

We now turn to the interesting for us case $1/a_{\1\2}\ll \{\sqrt{n_\1},\sqrt{n_\2}\} \ll \{1/a_{\1\1},1/a_{\2\2}\}$ where the interspecies interaction is weakly attractive and intraspecies ones are weakly repulsive. Let us introduce an auxiliary energy parameter $\Delta$ and a new set of coupling constants defined by $\tilde{g}_{\sigma\sigma'}=4\pi/\ln(\epsilon_{\sigma\sigma'}/\Delta)$. We choose $\Delta$ such that $\tilde{g}^2_{\1\2}=\tilde{g}_{\1\1}\tilde{g}_{\2\2}$ or, explicitly, $\Delta = \sqrt{\epsilon_{\1\2}\sqrt{\epsilon_{\1\1}\epsilon_{\2\2}}}\exp [-\ln^2(\epsilon_{\1\1}/\epsilon_{\2\2})/4\ln(\epsilon_{\1\1}\epsilon_{\2\2}/\epsilon_{\1\2}^2)].$ Then we substitute the expansion $g_{\sigma\sigma'}\approx \tilde{g}_{\sigma\sigma'}[1+\tilde{g}_{\sigma\sigma'}\ln(\kappa^2/\Delta)/4\pi+...]$ into Eq.~(\ref{BogKappa}) and keep terms up to second order in the new small parameters $\tilde{g}_{\sigma\sigma'}$. The energy density then reads
\begin{eqnarray}\label{BogDroplet}
E_{\rm 2D}&=&\frac{1}{2}(\tilde{g}_{\1\1}^{1/2}n_\1-\tilde{g}_{\2\2}^{1/2}n_\2)^2\nonumber \\
&&\hspace{-1.cm}+\frac{1}{8\pi}(\tilde{g}_{\1\1}n_\1+\tilde{g}_{\2\2}n_\2)^2\ln\frac{(\tilde{g}_{\1\1}n_\1+\tilde{g}_{\2\2}n_\2)\sqrt{e}}{\Delta}.\hspace{0.6cm}
\end{eqnarray}
Properties of the liquid phase in free space are obtained by minimizing the grand potential density $E_{\rm 2D}-\mu_\1 n_\1-\mu_\2 n_\2$ and by requiring that its value be zero (zero pressure). Explicitly, $E_{\rm 2D}-\sum_\sigma (\partial E_{\rm 2D}/\partial n_\sigma)n_\sigma =0$. One can show that possible values of $n_\1$ and $n_\2$  are close to the line $n_\1/n_\2=\sqrt{\tilde{g}_{\2\2}/\tilde{g}_{\1\1}}$ where the dominant first-order term in Eq.~(\ref{BogDroplet}) vanishes. Particularly, for $n = n_\1= n_\2\sqrt{\tilde{g}_{\2\2}/\tilde{g}_{\1\1}}$ Eq.~(\ref{BogDroplet}) reduces to the form $\propto \tilde{g}^2n^2[\ln(n/n_0)-1]$, where $n_0\sim \Delta/|\tilde{g}|$ is the equilibrium density at which the grand potential vanishes or, equivalently, the energy per particle $\propto E_{\rm 2D}/n$ reaches its minimum as a function of $n$. We {\it a posteriori} verify that $\kappa^2/\Delta$ is not exponentially large and, therefore, the small parameters $g$ and $\tilde{g}$ are equivalent.


In the symmetric case $a_{\1\1}=a_{\2\2}=a$ and $n_\1 = n_\2 = n$, one has $\Delta=\sqrt{\epsilon_{\1\2}\epsilon_{\1\1}}$, the energy density simplifies to
\begin{equation}\label{PopovEn}
E_{\rm 2D}=\frac{8\pi n^2}{\ln^2(a_{\1\2}/a)}[\ln(n/n_0)-1],
\end{equation}
and the equilibrium density of each component reads
\begin{equation}\label{PopovDensity}
n_0=\frac{e^{-2\gamma-3/2}}{2\pi}\frac{\ln(a_{\1\2}/a)}{aa_{\1\2}}.
\end{equation}
The knowledge of the equation of state (\ref{PopovEn}) permits us to find the spinodal point. Defined by the condition $\partial^2 E_{\rm 2D}/\partial n^2=0$ it is located at $n= e^{-1/2} n_0 \approx 0.61 n_0$. The mixture is thus metastable for $0.61n_0<n<n_0$. Note that since $1/\ln(a_{\1\2}/a)\ll 1$ the parameter $na^2\propto (a/a_{\1\2})\ln(a_{\1\2}/a)$ is exponentially small. We are thus dealing  with an extremely dilute liquid qualitatively different from usual liquids where $na^2\sim 1$.

\begin{center}
\begin{figure}[ht]
\vskip 0 pt \includegraphics[clip,width=1\columnwidth]{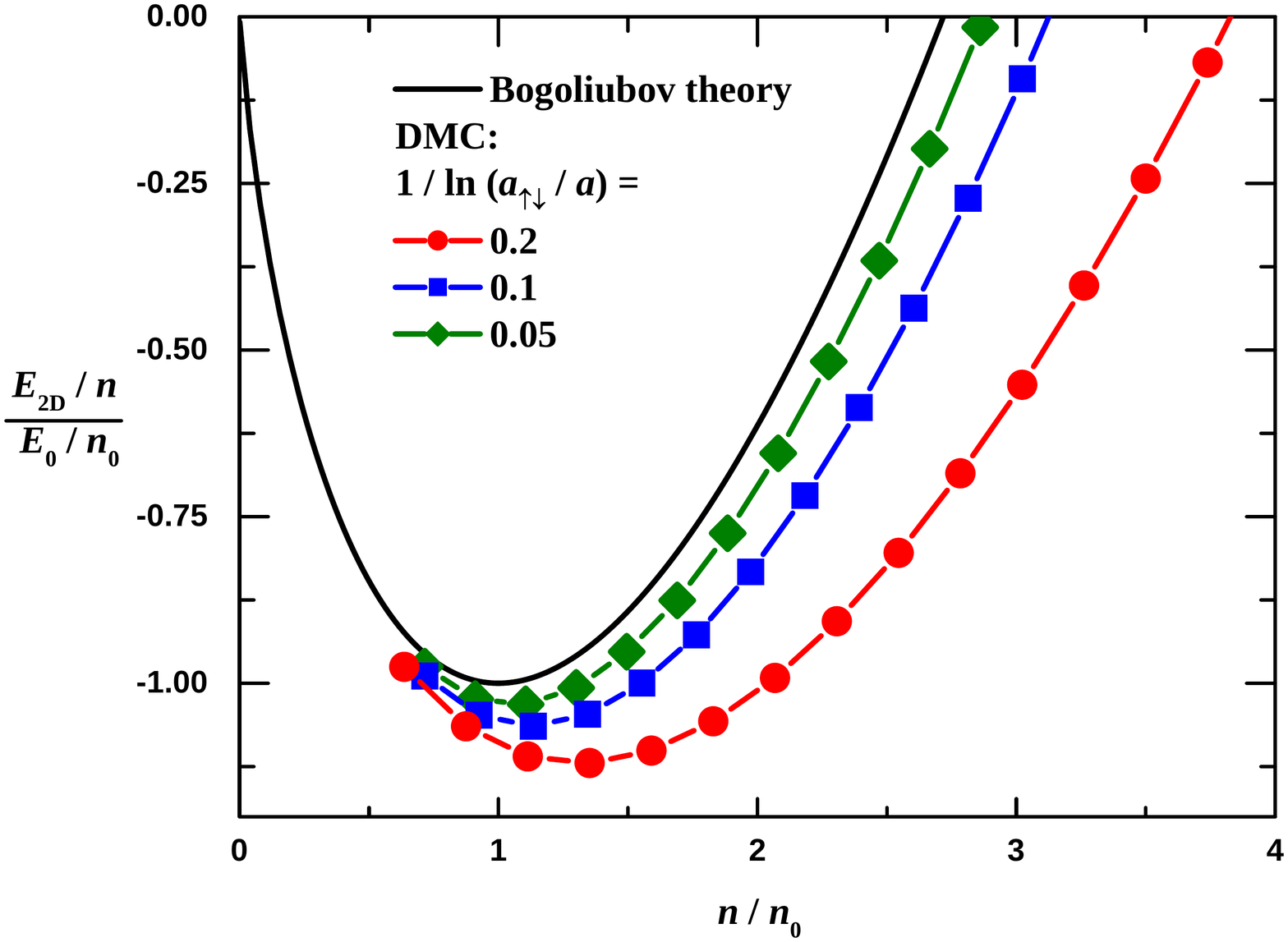}
\caption{
The energy per particle $E_{\rm 2D}/2n$ versus $n$ for the two-dimensional mixture with $a_{\1\1}=a_{\2\2}=a$ and $n_\1=n_\2=n$. We rescale the vertical and horizontal axes respectively by $E_{0}/2n_0=|E_{\rm 2D}(n_0)|/2n_0$ and $n_0$ calculated in the Bogoliubov approximation [Eqs.~(\ref{PopovEn}-\ref{PopovDensity})].
The solid black line is the result of Eq.~(\ref{PopovEn}) and the scattered data are the DMC results for $1/\ln(a_{\1\2}/a)=0.2$ (red, circles), 0.1 (blue, squares), and 0.05 (green, diamonds) corresponding to $n_0a^2=3.8\times 10^{-4}$, $5.1\times 10^{-6}$, and $4.6\times 10^{-10}$, respectively. The interspecies (intraspecies) interactions are modeled by square wells (soft disks) with the range $R_0$ fixed by $nR_0^2=5\times 10^{-3}$ and with the depths (heights) adjusted in order to obtain the desired values of $a_{\sigma\sigma'}$. As a universality check we have significantly reduced $R_0$ and found that the energy changes at most by the symbol size.
}
\label{Fig:2D}
\end{figure}
\par\end{center}

In order to check the universality of our theory and importance of higher-order corrections we perform diffusion Monte Carlo (DMC) calculations for the symmetric mixture ($n=n_\1=n_\2$ and $a=a_{\1\1}=a_{\2\2}$) for various densities and interaction potentials. The DMC method gives the ground-state energy exactly and it has been applied to the one-component two-dimensional Bose gas in Ref.~\cite{Pilati05}. In our case the convergence is enhanced by using the guiding wave function in the Jastrow pair-product form where we match the inter- and intraspecies two-body scattering solutions at short distances with the long-range phononic behavior at large distances \cite{ReattoChester67}. The calculations are performed in a finite box with periodic boundary conditions and the results are extrapolated to the thermodynamic limit \cite{SM,RemSpinodal}. In Fig.~\ref{Fig:2D} we present the density dependence of the energy per particle. As expected, with decreasing $1/\ln(a_{\1\2}/a)$ the numerical results converge towards our theory and the rate of this convergence is consistent with the scaling $n^2/\ln^3(a_{\1\2}/a)$ for the next-order correction to Eq.~(\ref{PopovEn}).

Let us now comment on the applicability of the above results to quasi-two-dimensional atomic mixtures. The passage from three-dimensional scattering parameters to two-dimensional ones is realized by using the formula $\epsilon_{\sigma\sigma'}=(B/\pi l_0^2)\exp[\sqrt{2\pi}l_0/a_{\sigma\sigma'}^{(3D)}]$ \cite{PetrovShlyapnikov}, where $a_{\sigma\sigma'}^{(3D)}$ are the three-dimensional scattering lengths, $l_0$ is the oscillator length in the confinement direction, and $B\approx 0.9$. In particular, in the symmetric case, $a_{\1\1}^{(3D)}=a_{\2\2}^{(3D)}=a^{(3D)}$, the equilibrium densities of the components equal
\begin{equation}\label{Symmetricn0Quasi2D}
n_0=\frac{B[1/a^{(3D)}-1/a_{\1\2}^{(3D)}]}{4(2\pi e)^{3/2}l_0}e^{\sqrt{\pi/2}[l_0/a_{\1\2}^{(3D)}+l_0/a^{(3D)}]}.
\end{equation}
The weakly interacting regime in this case is ensured by the inequality $\ln (a_{\1\2}/a)=\sqrt{\pi/2}[l_0/a^{(3D)}-l_0/a_{\1\2}^{(3D)}]\gg 1$ and the requirement that typical transverse energies be much smaller than $1/l_0^2$ (two-dimensional regime) practically reduces to $-[l_0/a_{\1\2}^{(3D)}+l_0/a^{(3D)}]\gg 1$. We can rewrite these two conditions as $0<-a_{\1\2}^{(3D)}<a^{(3D)}\ll l_0$. Note that a three-dimensional mixture satisfying $0<-a_{\1\2}^{(3D)}<a^{(3D)}$ is in the stable gas phase since the interspecies attraction is too weak. We thus find a curious fact that by introducing the confinement the mixture becomes liquid. The nonsymmetric case is analysed in the same fashion and we finally note that suitable combinations of $a_{\sigma\sigma'}^{(3D)}$ are available for hyperfine components $F=1,m_F=-1$ and $F=1,m_F=0$ of $^{39}$K \cite{Simoni,Lysebo}.

Let us now discuss finite-size droplets of the liquid. The derivation of the corresponding Gross-Pitaevskii equation follows the same path as in the three-dimensional case \cite{PetrovDroplet}. In short, the length scale on which the droplet profile changes is of order $\xi\sim 1/\sqrt{|\mu|}$, where the chemical potential $\mu\sim -n/\ln^2(a_{\1\2}/a)$ [see Eq.~(\ref{PopovEn})]. On the other hand, excitations mostly contributing to the second-order terms in Eqs.~(\ref{BogIntermediate}) and (\ref{BogKappa}) belong to the upper Bogoliubov branch and have wavelengths $\sim 1/c_+ \propto 1/\sqrt{n/|\ln(a_{\1\2}/a)|}\ll \xi$. This separation of scales means that in the effective theory for fields with momenta $k\sim \sqrt{|\mu|}$ the effect of higher-momentum modes is just a local density-dependent term and one can write the energy density functional as
\begin{equation}\label{EnergyDensity}
\varepsilon(\psi,\psi^*)=|\nabla \psi|^2+\frac{8\pi|\psi|^4}{\ln^2(a_{\1\2}/a)}\ln\frac{|\psi|^2}{en_0}.
\end{equation}
Here the complex field $\psi(\boldsymbol{\rho},t)$ satisfies the normalization condition $N=\int |\psi(\boldsymbol{\rho},t)|^2 d^2\rho$, where $N$ is the number of particles in each component (we consider the symmetric case). The Gross-Pitaevskii equation for $\psi$ reads
\begin{equation}\label{GP}
i\dot{\psi}=-\frac{\nabla^2}{2}\psi+\frac{8\pi}{\ln^2(a_{\1\2}/a)}\ln\left(\frac{|\psi|^2}{\sqrt{e}n_0}\right)|\psi|^2\psi
\end{equation}
and the stationary one is obtained from Eq.~(\ref{GP}) by substituting $\psi(\boldsymbol{\rho},t)=\psi(\boldsymbol{\rho})e^{-i\mu t}$ [for uniform liquid $\mu=\mu_0 =-4\pi n_0/\ln^2(a_{\1\2}/a)$]. The dimensional analysis of Eq.~(\ref{GP}) shows that the typical length scale on which $\psi$ changes is indeed $\xi$. If $\psi$ is real and depends only on one coordinate, say $x$, this type of equation (with no explicit spatial dependence of coefficients) maps to the classical problem of a particle moving in time $x$ and coordinate $\psi$ \cite{Soliton}. We will discuss it in more detail in the one-dimensional case. Here we mention that the surface tension (the energy per unit length of the liquid-vacuum interface) $\sigma=\int dx[\varepsilon(\psi,\psi*)-2\mu_0 |\psi|^2]= I\sqrt{\pi}(2 n_0)^{3/2}/\ln(a_{\1\2}/a)$, where $I=\int_0^1 dz \sqrt{1-z+z\ln z}\approx 0.42$. This quantity is useful for calculating finite-size corrections to droplet's energy and the spectrum of its surface modes (see, for example, \cite{Bulgac}). Note that such droplets with almost uniform bulk density qualitatively differ from exponentially small and dense many-body bound states of attractive two-dimensional scalar bosons stabilized by the increased kinetic energy associated with their nonuniform shape \cite{HammerSon}.


We now turn to the one-dimensional case where the weakly interacting regime requires $|g_{\sigma\sigma'}|/n\ll 1$ \cite{Olshanii1998}. Strictly speaking, there is no condensate in one dimension, but it is now well understood that the energy of a weakly interacting Bose gas is correctly predicted by the Bogoliubov theory which assumes condensate \cite{LiebLiniger,Popov1971}. In this way we obtain the energy density in the form of Eq.~(\ref{BogIntermediate}) where no cutoff is necessary, and the integration over momentum results in
\begin{equation}\label{Bog1D}
E_{\rm 1D}=\frac{1}{2}\sum_{\sigma\sigma'}g_{\sigma\sigma'}n_\sigma n_{\sigma'}-\frac{2}{3\pi}\sum_{\pm}c_{\pm}^3,
\end{equation}
where $c_{\pm}$ are given by Eq.~(\ref{cpm}).

Let us introduce $\delta g = g_{\1\2}+\sqrt{g_{\1\1}g_{\2\2}}$ and discuss the regime of repulsive intra- and attractive interspecies interactions close to the mean-field collapse instability point such that $0<\delta g\ll g=\sqrt{g_{\1\1}g_{\2\2}}$. In this regime Eq.~(\ref{Bog1D}) can be rewritten as
\begin{eqnarray}\label{BogDroplet1D}
E_{\rm 1D}&=&\frac{(g_{\1\1}^{1/2}n_\1-g_{\2\2}^{1/2}n_\2)^2}{2}+\frac{g\delta g(g_{\2\2}^{1/2}n_\1+g_{\1\1}^{1/2}n_\2)^2}{(g_{\1\1}+g_{\2\2})^2}\nonumber \\
&&\hspace{-0.cm}-\frac{2}{3\pi}(g_{\1\1}n_\1+g_{\2\2}n_\2)^{3/2}.\hspace{0.6cm}
\end{eqnarray}
Similarly to the higher-dimensional cases we assume $n=n_\uparrow=n_\downarrow \sqrt{g_{\downarrow\downarrow}/g_{\uparrow\uparrow}}$. Then the structure of the energy density is $E \propto \delta g n^2-(gn)^{3/2}$ characterized by the existence of the liquid state with equilibrium density $n_0\sim g^3/\delta g^2$. Note that at this density $g/n\sim (\delta g/g)^2 \ll 1$, i.e., the system is weakly interacting. Counterintuitively, liquid appears for $\delta g>0$ in the regime where the mixture is on average repulsive and where one would expect a stable gas phase. It is thus the attractive beyond-mean-field term that liquefies it.

The quantitative analysis of the droplet properties in the one-dimensional case as well as the derivation of the corresponding Gross-Pitaevskii equation goes along the same lines as in the higher-dimensional cases. In particular, in the symmetric case $g=g_{\1\1}=g_{\2\2}$ and $n=n_\1=n_\2$ Eq.~(\ref{BogDroplet1D}) becomes
\begin{equation}\label{Bog1Dsymm}
E_{\rm 1D}=\delta g n^2-4\sqrt{2}(gn)^{3/2}/3\pi,
\end{equation}
the equilibrium density reads
\begin{equation}\label{EqDensity1Dsymm}
n_0=8g^3/(9\pi^2\delta g^2)
\end{equation}
and the corresponding chemical potential equals $\mu_0=-\delta g n_0/2$. The spinodal point is at $n=9n_0/16\approx 0.56 n_0$. In Fig.~\ref{Fig:1D} we compare the prediction of Eq.~(\ref{Bog1Dsymm}) valid in the limit $\delta g/g\rightarrow 0$ with our DMC results obtained also for the symmetric mixture with delta-function interactions but at finite values of $\delta g/g$. Our numerical procedure is similar to the one used in the one-dimensional one-component case \cite{1Dcorrelations}. The rate of convergence towards Eq.~(\ref{Bog1Dsymm}) indicates that the expansion of the energy continues in integer powers of $\sqrt{g/n}\propto \delta g/g$.

\begin{center}
\begin{figure}[ht]
\vskip 0 pt \includegraphics[clip,width=1\columnwidth]{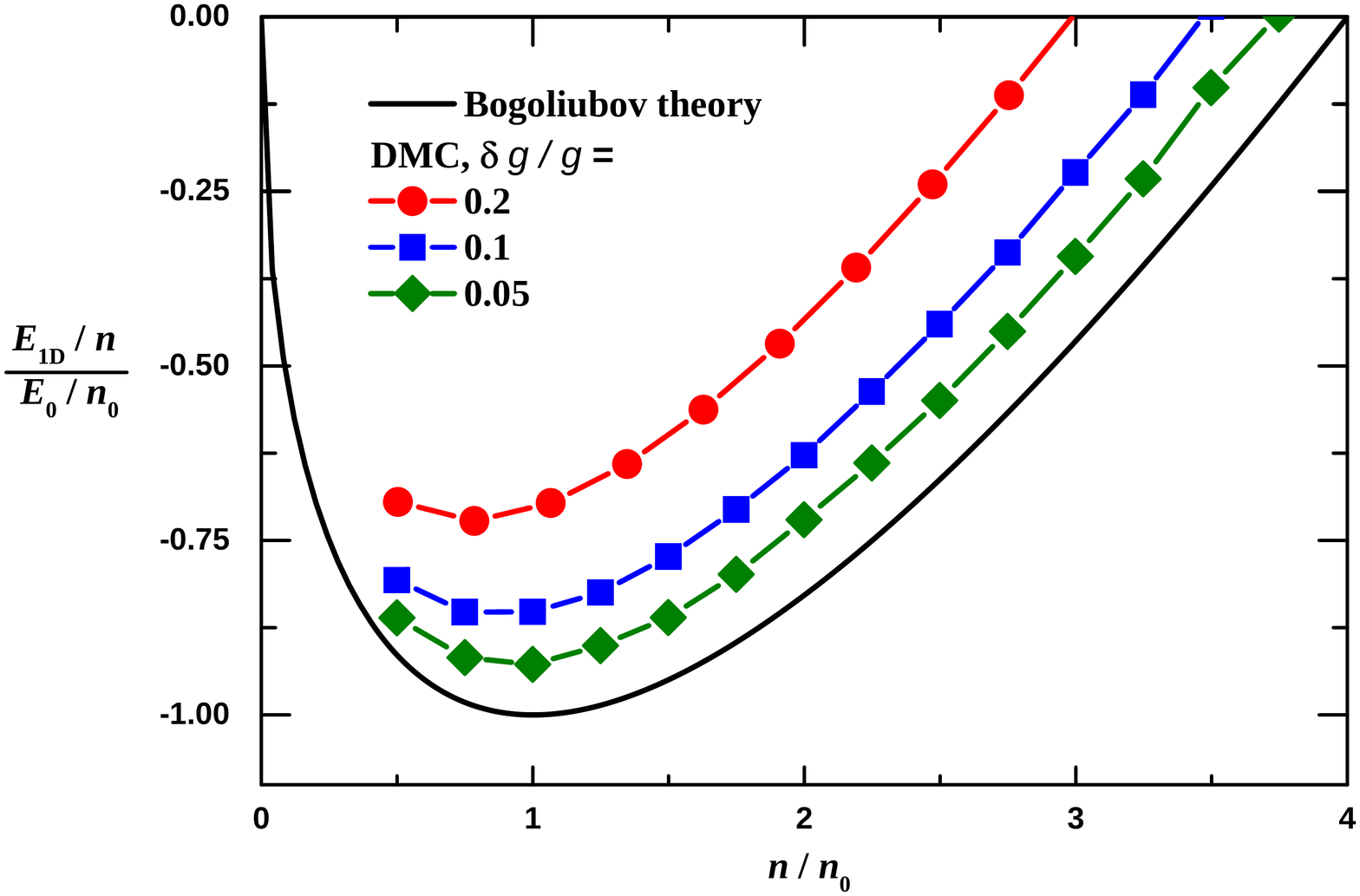}
\caption{The energy per particle $E_{\rm 1D}/2n$ versus $n$ for the symmetric one-dimensional mixture with delta-function interactions. The vertical and horizontal axes are rescaled respectively by $E_{0}/2n_0=|E_{\rm 1D}(n_0)|/2n_0$ and $n_0$ given from Eqs.~(\ref{Bog1Dsymm}-\ref{EqDensity1Dsymm}). The solid black line is given by Eq.~(\ref{Bog1Dsymm}), exact for $\delta g/g\rightarrow 0$, and the scattered data are the DMC results for  $\delta g/g=0.2$ (red circles), 0.1 (blue squares), and 0.05 (green diamonds).
}
\label{Fig:1D}
\end{figure}
\par\end{center}

The Gross-Pitaevskii equation for the droplet reads
\begin{equation}\label{GP1D}
i\dot{\psi}=-\psi''_{xx}/2+\delta g|\psi|^2\psi-(\sqrt{2}/\pi)g^{3/2}|\psi|\psi,
\end{equation}
where $\psi(x,t)$ satisfies $N=\int |\psi(x,t)|^2 dx$. It turns out that the droplet exists for any $\mu_0<\mu<0$ (which translates to any $N$) and its shape can be found analytically. For real $\psi$ Eq.~(\ref{GP1D}) can be written in the form $\psi''_{xx}=-V'_\psi (\psi)$, where $V(\psi)=-\delta g \psi^4/2+(2\sqrt{2}/3\pi)g^{3/2}\psi^3+\mu \psi^2$. This equation describes the trajectory of a classical particle in time $x$ with coordinate $\psi$ \cite{Soliton}. Once integrated, it reads $d\psi/\sqrt{-2V(\psi)}=dx$. The second integration gives the shape of the droplet
\begin{equation}\label{DropletShape1D}
\psi(x,t)=\frac{\sqrt{n_0}e^{-i\mu t}\mu/\mu_0}{1+\sqrt{1-\mu/\mu_0}\cosh(\sqrt{-2\mu}x)}
\end{equation}
containing $N=2\sqrt{\frac{n_0}{\delta g}}\left[\ln\frac{1+\sqrt{\mu/\mu_0}}{\sqrt{1-\mu/\mu_0}} - \sqrt{\frac{\mu}{\mu_0}}\right]$
particles of each component. Note that in contrast to the usual single-soliton solution of the one-dimensional Schr\"odinger equation with attractive cubic nonlinearity \cite{ZakharovShabat} our droplet has a flat bulk region for $\mu\approx \mu_0$. We also note that in this case the typical inverse length on which $\psi$ changes is of order $\sqrt{\delta g n_0}$ which is much smaller than the typical momentum $\sim \sqrt{g n_0}$ contributing to the last (beyond-mean-field) term in Eq.~(\ref{Bog1Dsymm}). This justifies the low-energy theory (\ref{GP1D}). For $\mu/\mu_0\ll 1$ the size of the droplet increases and its density decreases with decreasing $|\mu|$. For the validity of (\ref{GP1D}) in this case we need $\mu/\mu_0\gg \delta g/g$ or, equivalently, $N\gg 1$.

In conclusion, weakly interacting low-dimensional Bose-Bose mixtures manifest themselves as promising candidates for studying liquid phases in the ultracold ultradilute regime and associated beyond-mean-field effects. We find that in the two-dimensional case the liquid phase is formed whenever the intraspecies interactions are repulsive and the interspecies one is attractive. This differs from the three-dimensional case where $|g_{\1\2}|$ should be larger than $\sqrt{g_{\1\1} g_{\2\2}}$. Remarkably, the one-dimensional mixture liquefies for $|g_{\1\2}|<\sqrt{g_{\1\1} g_{\2\2}}$; this effect is completely missed by the mean-field approximation. Interestingly, one-dimensional droplets can be described analytically and it is tempting to study their dynamical and transport properties. In any dimension the almost complete cancellation of the first-order terms in the energy functional of the liquid gives one an opportunity to test higher-order terms and their universality. It is then relevant, although theoretically challenging, to go beyond the Bogoliubov approximation as it has been done in the scalar two-dimensional case \cite{MoraCastin}.

We acknowledge support by the IFRAF Institute. The research leading to these results received funding from the European Research Council (FR7/2007-2013 Grant Agreement No. 341197) and the MICINN (Spain) Grant No. FIS2014-56257-C2-1-P. The Barcelona Supercomputing Center (The Spanish National Supercomputing Center - Centro Nacional de Supercomputaci\'on) is acknowledged for the provided computational facilities.


\newpage

\renewcommand{\theequation}{S\arabic{equation}}
\renewcommand{\thefigure}{S\arabic{figure}}

\setcounter{equation}{0}
\setcounter{figure}{0}

\begin{widetext}

\centerline{\large{\bf SUPPLEMENTAL MATERIAL}}
\vspace{1cm}

The diffusion Monte Carlo (DMC) technique provides the exact ground-state energy when 
({\it i}) the simulation time goes to infinity, 
({\it ii}) time step goes to zero, and  
({\it iii}) population size goes to infinity. Deviations from these limits introduce statistical and systematic errors which we control and minimize according to a given accuracy goal. The convergence towards limits ({\it i})-({\it iii}) depends on the choice of the guiding wave function which we discuss in Sec.~\ref{Sec:wf}. The statistical error is reduced by making the simulation series large enough (the statistical error comes from a finite simulation time and is estimated by a standard block-averaging procedure). As for the systematic errors, to improve the convergence we use a quadratic time step algorithm with time step $\Delta t = 0.01$ in units where the mass $m$, mean interparticle separation $n^{-1/d}$, and $\hbar$ are equal to 1. We have verified that the diffusion algorithm without branching for this time step recovers the variational energy within the accuracy goal. All simulations have been performed with the population size of 1000 walkers which we find to be sufficient. The calculations are done in a box with periodic boundary conditions for various particle numbers and we extrapolate the result to the thermodynamic limit. This procedure is well controlled since we know the finite-size correction in the Bogoliubov approximation (see Sec.~\ref{Sec:finite size}). In the two-dimensional case we choose sufficiently short-range interaction potentials in order to claim that the results are valid in the zero-range limit (see Sec.~\ref{Sec:finite range}). In one dimension we work directly with zero-range potentials.

\section{Guiding wave function\label{Sec:wf}}

We chose the guiding wave function in the pair-product form
\begin{eqnarray}
\Psi_T(
{\bf r}_1^\1, \cdots, {\bf r}_{N_\1}^\1,
{\bf r}_1^\2, \cdots, {\bf r}_{N_\2}^\2
)
=
\prod\limits_{i<j}^{N_\1} f_{\1\1}(|{\bf r}_i^\1 - {\bf r}_j^\1|)
\prod\limits_{i<j}^{N_\2} f_{\2\2}(|{\bf r}_i^\2 - {\bf r}_j^\2|)
\prod\limits_{i=1}^{N_\1}
\prod\limits_{j=1}^{N_\2} f_{\1\2}(|{\bf r}_i^\1 - {\bf r}_j^\2|)
\label{Eq:wftrial}
\end{eqnarray}
and  seek to incorporate as much physical information into the Jastrow terms $f_{\sigma\sigma'}(r)$ as possible.
Reatto and Chester\cite{ReattoChester67S} showed by using hydrodynamic approach that the ``phononic'' long-range part of a single component many-body wave function $\Psi({\bf r}_1, \cdots, {\bf r}_N)$ can be written in a Jastrow pair-product form
\begin{eqnarray}
\Psi({\bf r}_1, \cdots, {\bf r}_N)
= \exp\left[- \frac{1}{2} \sum\limits_{i<j} \chi(|{\bf r}_i - {\bf r}_j|)\right],
\label{Eq:wfphonons}
\end{eqnarray}
where asymptotic long-range decay is $\chi(r) = mc /(\pi^2 n \hbar r^2) \propto 1/r^2$ in three dimensions and $\chi(r) \propto 1/r$ in two dimensions.
Here $n$ is the density and $c$ is the speed of sound, corresponding to the long-wavelength phonons.
We chose a form similar to that of Eq.~(\ref{Eq:wfphonons}) for describing the long-range part of $f_{\sigma\sigma'}(r)$ in~(\ref{Eq:wftrial}).

When two particles come close to each other, the dominant physical process is the scattering between those two particles.
We chose the short range part of $f_{\sigma\sigma'}(r)$ as a solution of the two-body scattering problem for the corresponding interaction potential $U_{\sigma\sigma'}(r)$.

\subsection{Two dimensions\label{Sec:wf2D}}
In the two-dimensional case, the intraspecies interactions are modeled by soft disks (SD)
\begin{eqnarray}
U_{\1\1}(r)
= U_{\2\2}(r)
=
\begin{cases}
    U^{SD}_0,& \text{if } r\leq R_0\\
    0,              & \text{otherwise}
\end{cases}
\label{Eq:V:SD}
\end{eqnarray}
and the interspecies interactions by square wells (SW)
\begin{eqnarray}
U_{\1\2}(r)
=
\begin{cases}
    -U^{SW}_0,& \text{if } r\leq R_0\\
    0,              & \text{otherwise}
\end{cases}
\label{Eq:V:SW}
\end{eqnarray}
with the same range $R_0$.
The height $U^{SD}_0 > 0$ of the soft disk and the depth $U^{SW}_0 > 0$ of the square well are adjusted in order to obtain the desired values of the $s$-wave scattering length $a_{\sigma\sigma'}$.
We also considered hard disks (HD), obtained from soft disks in the limit of an infinite height of the interaction potential $U^{SD}_0 \to \infty$, in which the $s$-wave scattering length $a$ corresponds to the diameter of the hard disk, $a = R_0$.

The following Jastrow terms are used in two dimensions:
\begin{itemize}
\item for the soft-disk potential
\begin{eqnarray}
f^{SD}_{\1\1}(r)
= f^{SD}_{\2\2}(r)
=
\begin{cases}
A I_0(\kappa r),& \text{if } r\leq R_0\\
B \ln(r / a),& \text{if } R_0 < r \leq R_{par}\\
C \exp(-D/r + E/r^2), & \text{if } R_{par} < r \leq L/2\\
1,              & \text{if } r>L/2
\end{cases}
\label{Eq:wf:SD}
\end{eqnarray}
%
\item for the hard-disk potential
\begin{eqnarray}
f^{HD}_{\1\1}(r)
= f^{HD}_{\2\2}(r)
=
\begin{cases}
0,& \text{if } r\leq R_0\\
A \ln(r / a),& \text{if } R_0 < r \leq R_{par}\\
B \exp(-C/r + D/r^2), & \text{if } R_{par} < r \leq L/2\\
1,              & \text{if } r>L/2
\end{cases}
\label{Eq:wf:HD}
\end{eqnarray}
%
\item for the square-well potential
\begin{eqnarray}
f^{SW}_{\1\2}(r)
=
\begin{cases}
A J_0(\kappa^{\1\2} r),& \text{if } r\leq R_0\\
B \ln(r / a_{\1\2}),& \text{if } R_0 < r \leq R^{\1\2}_{par}\\
C \exp(-D/r + E/r^2), & \text{if } R^{\1\2}_{par} < r \leq L/2\\
1,              & \text{if } r>L/2
\end{cases}
\label{Eq:wf:SW}
\end{eqnarray}
\end{itemize}
Here, $I_0(r)$ and $J_0(r)$ are modified Bessel function of the first kind. The characteristic momenta $\kappa$ and $\kappa^{\1\2}$ are defined by the height (depth) of the interaction potential according to
$\kappa = \sqrt{m U^{SD}_0} / \hbar$
and
$\kappa^{\1\2} = \sqrt{m U^{SW}_0}/ \hbar$.
The short-range part, $r<R_{par}$ and $r<R_{par}^{\1\2}$, corresponds to the zero-energy scattering solution on interaction potentials~(\ref{Eq:V:SD}) and~(\ref{Eq:V:SW}).
The long-range part, $r>R_{par}$ and $r>R_{par}^{\1\2}$, has the phononic asymptotic~(\ref{Eq:wfphonons}).
Coefficients $A$, $B$, $C$, $D$, $E$ are fixed by the conditions of the continuity of the function itself, $f(r)$, its first derivative, $f'(r)$, and by the periodic boundary conditions which are satisfied by imposing zero derivative at the half size of the box, $f'(L/2) = 0$.
The variational parameters $R_{par}$ and $R_{par}^{\1\2}$ are optimized by minimizing the variational energy.
For the repulsive interactions (SD and HD), parameter $R_{par}$ corresponds to the matching distance between the two-body scattering and the phononic regimes.
For attractive SW interaction, parameter $R_{par}^{\1\2}$ effectively changes the value of $f(r=0)$ and physically describes how strongly is localized a pair of two particles in the many-body system.

\subsection{One dimension\label{Sec:wf1D}}

In the one-dimensional case we perform simulations directly for the $\delta$-pseudopotential~\cite{1DcorrelationsS} thus avoiding any finite-range bias.
This can be done by imposing Bethe-Peierls boundary condition on the many-body wave function 
\begin{eqnarray}
\frac{d}{dr}[f^{\sigma\sigma'}(r)]_{r=0} = - \frac{1}{a_{\sigma\sigma'}}[f^{\sigma\sigma'}(r)]_{r=0}.
\label{Eq:Bethe-Peierls}
\end{eqnarray}
The $\delta$-pseudopotential acts only at the contact point, $|r|=0$, while for any finite separation between two particles a good choice for the short-range part of the Jastrow terms is a plane wave (repulsive interaction) or a decaying exponent (attractive interaction).

We take the long-range part of Jastrow terms from the hydrodynamic expression, Eq.~(\ref{Eq:wfphonons}).
The presence of phonons in 1D induce slowly-decaying quantum correlations \cite{ReattoChester67S} between particles,
$\chi(r) = - (2/K_L) \ln[\sin|\pi r / L|]$, where $K_L = \pi \hbar n / (mc)$ is the Luttinger parameter.
As a result, instead of exponential Jastrow terms in higher dimensions, here the decay is instead of a power-law type,
$f(r) = |\sin|\pi r / L||^{1/K_L}$.

The following Jastrow terms are used in two dimensions:
\begin{itemize}
\item for the repulsive intraspecies $\delta$-pseudopotenial
\begin{eqnarray}
f_{\1\1}(r)
= f_{\2\2}(r)
=
\begin{cases}
A \cos(k|r-B|), & \text{if } |r| \leq R_{par}\\
|\sin(|\pi r /L|)|^{1/K_{par}}, & \text{if } R_{par} < |r| \leq L/2\\
1,              & \text{if } |r|>L/2
\end{cases}
\label{Eq:pseudopotential}
\end{eqnarray}
%
\item for the attractive interspecies $\delta$-pseudopotenial
\begin{eqnarray}
f_{\1\2}(r)
=
\begin{cases}
A \exp(-|r|/a_{\1\2}), & \text{if } |r| \leq R_{par}^{\1\2}\\
|\sin(\pi |r| /L)|^{1/K_{par}^{\1\2}}, & \text{if } R_{par}^{\1\2} < |r| \leq L/2\\
1,              & \text{if } |r|>L/2
\end{cases}
\label{Eq:McGuire}
\end{eqnarray}
\end{itemize}
where coefficient $A$ is chosen according to the continuity condition at the matching point.
Periodic boundary conditions are automatically satisfied by using hydrodynamic/Luttinger-liquid tails. 
Coefficients $R_{par}$, $R_{par}^{\1\2}$, $K_{par}$ and , $K_{par}^{\1\2}$ are optimized by minimizing the variational energy.
We note that the exponential short-range part in Eq.~(\ref{Eq:McGuire}) reminds the exact wave function by McGuire for an attractive single-component Bose gas.



\section{Finite-size effects\label{Sec:finite size}}

As stated in the main text, the ground-state energy in the Bogoliubov approximation is given by
\begin{equation}
\label{BogIntermediateS}
E = \frac{1}{2}\sum_{\sigma\sigma'}g_{\sigma\sigma'}n_\sigma n_{\sigma'}+\frac{1}{2}\sum_\pm \sum_{|{\bf k}|<\kappa}[E_\pm (k)-k^2/2-c_\pm^2]\;.
\end{equation}
In a box of size $L^d$ with periodic boundary conditions the summation is performed over momenta ${\bf k} = 2\pi {\bf n} / L$, where vector ${\bf n}$ is a list of $d$ integers. In the thermodynamic limit we replace the summation over ${\bf k}$ by integration. The finite-size correction originates from the discrete character of the lowest Bogoliubov modes and scales as a power of the ratio $\xi/L$, where $\xi=1/c_+$ is the healing length corresponding to the Bogoliubov $+$ mode.

\subsection{Two dimensions\label{Sec:finite size:2D}}

The ground-state energy density of the two-dimensional mixture with $a_{\1\1}=a_{\2\2}=a$ and $n_\1 = n_\2 = n$ including the leading-order finite-size correction reads
\begin{equation}\label{PopovEnS}
E_{\rm 2D}
=\frac{8\pi n^2}{\ln^2(a_{\1\2}/a)}[\ln(n/n_0)-1]
 + C \sqrt{\frac{2\pi}{\ln(a_{\1\2}/a)}}\frac{n^2}{N^{3/2}}\;,
\end{equation}
where
\begin{equation}\label{2DFiniteSize}
C = 2\pi \lim_{\alpha\rightarrow 0}\left[\sum_{m,n}\sqrt{m^2+n^2}e^{-\alpha(m^2+ n^2)}-\int |k| e^{-\alpha k^2} d^2k\right] \approx -1.438.
\end{equation}

\begin{center}
\begin{figure}[ht]
\vskip 0 pt \includegraphics[clip,width=0.6\columnwidth]{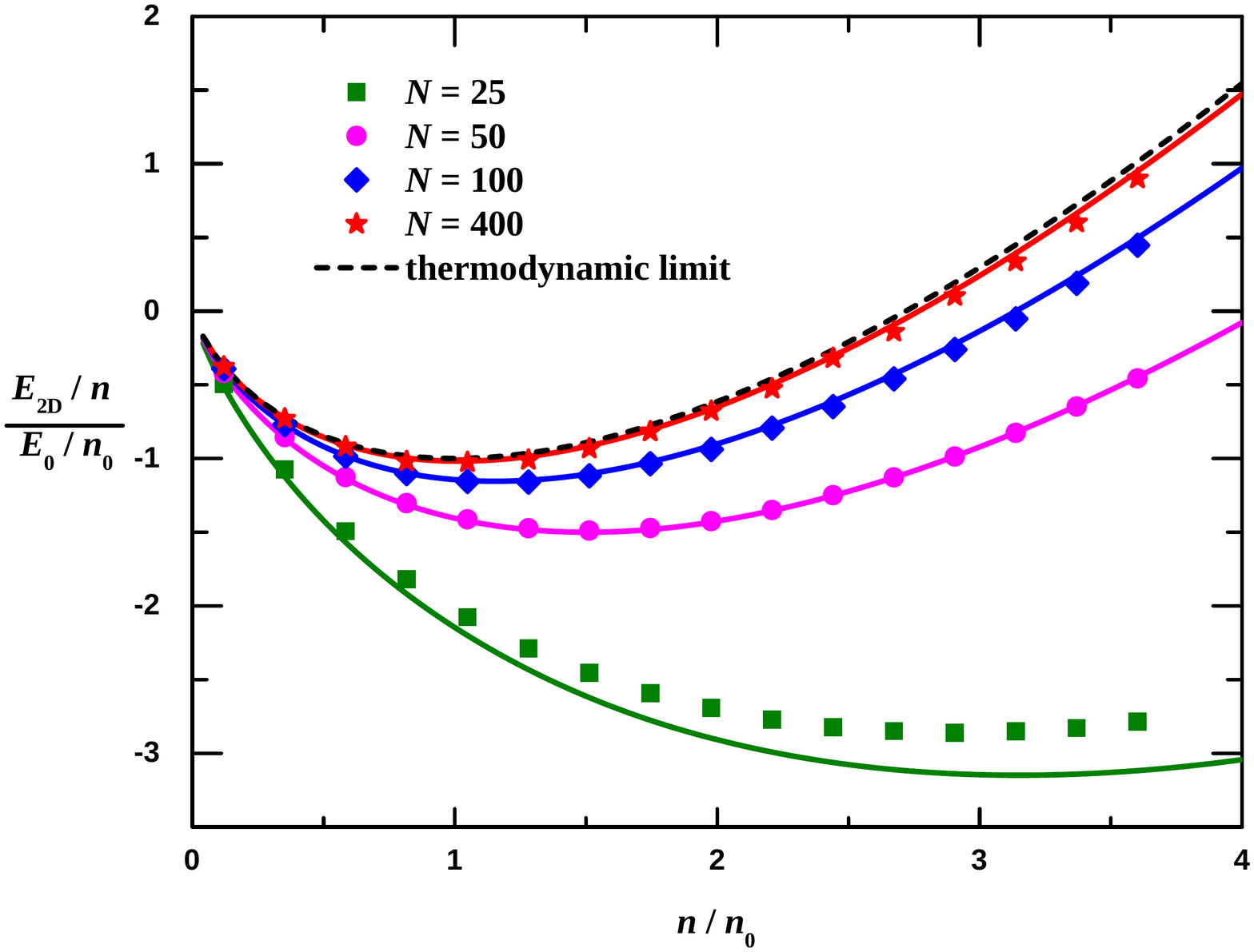}
\caption{
The energy per particle $E_{\rm 2D}/2n$ versus $n$ for the two-dimensional mixture with $a_{\1\1}=a_{\2\2}=a$ and $n_\1=n_\2=n$ with
$1/\ln(a_{\1\2}/a) = 0.01$ in a box with periodic boundary conditions for particle numbers in each component equal to $N=25$ (green squares), 50 (pink circles), 100 (blue diamonds), and 400 (red stars). Solid curves are predictions of the Bogoliubov theory with the leading-order finite-size correction, Eq.~(\ref{PopovEnS}). Dashed line is the Bogoliubov result in the thermodynamic limit.
}
\label{Fig:2Dfinitesize}
\end{figure}
\par\end{center}

Figure~\ref{Fig:2Dfinitesize} shows the energy per particle calculated for finite $N$ by using the DMC method (symbols) and the analytical result Eq.~(\ref{PopovEnS}) (solid lines) for $1/\ln(a_{\1\2}/a) = 0.01$. The relative deviation of finite-$N$ curves from the thermodynamic limit (dashed line) scales as $[\ln(a_{\1\2}/a)/N]^{3/2}\propto (\xi/L)^3$ and, for larger $1/\ln(a_{\1\2}/a)$, we need less atoms to reach the thermodynamic limit (within a given accuracy goal). Results presented in the main text are the extrapolation to this limit.

\subsection{One dimension\label{Sec:finite size:1D}}

In one dimension the ground-state energy density for the symmetric mixture ($g=g_{\1\1}=g_{\2\2}$ and $n=n_\1=n_\2$) including the leading finite-size correction reads
\begin{equation}
E_{\rm 1D}
=\delta g n^2
- \frac{4\sqrt{2}}{3\pi} (gn)^{3/2}
-\frac{\pi}{3\sqrt{2}} \frac{\sqrt{g} n^{5/2}}{N^2}
\;,
\label{Eq:Bog1Dsymm:finitesize}
\end{equation}
where the finite-size correction term is obtained by applying the Euler-Maclaurin formula to the one-dimensional sum over momenta in Eq.~(\ref{BogIntermediateS}). Figure~\ref{Fig:1Dfinitesize} shows the dependence of the energy per particle for various $N$ calculated by the DMC method (symbols) and by the Bogoliubov theory Eq.~(\ref{Eq:Bog1Dsymm:finitesize}) (solid lines) in the case $\delta g/g=0.05$. The relative deviation from the thermodynamic result (dashed line) scales as $(N\delta g/g)^{-2}\propto (\xi/L)^2$. Our DMC results reported in the main text are the extrapolation to $N\rightarrow \infty$.

\begin{center}
\begin{figure}[ht]
\vskip 0 pt \includegraphics[clip,width=0.6\columnwidth]{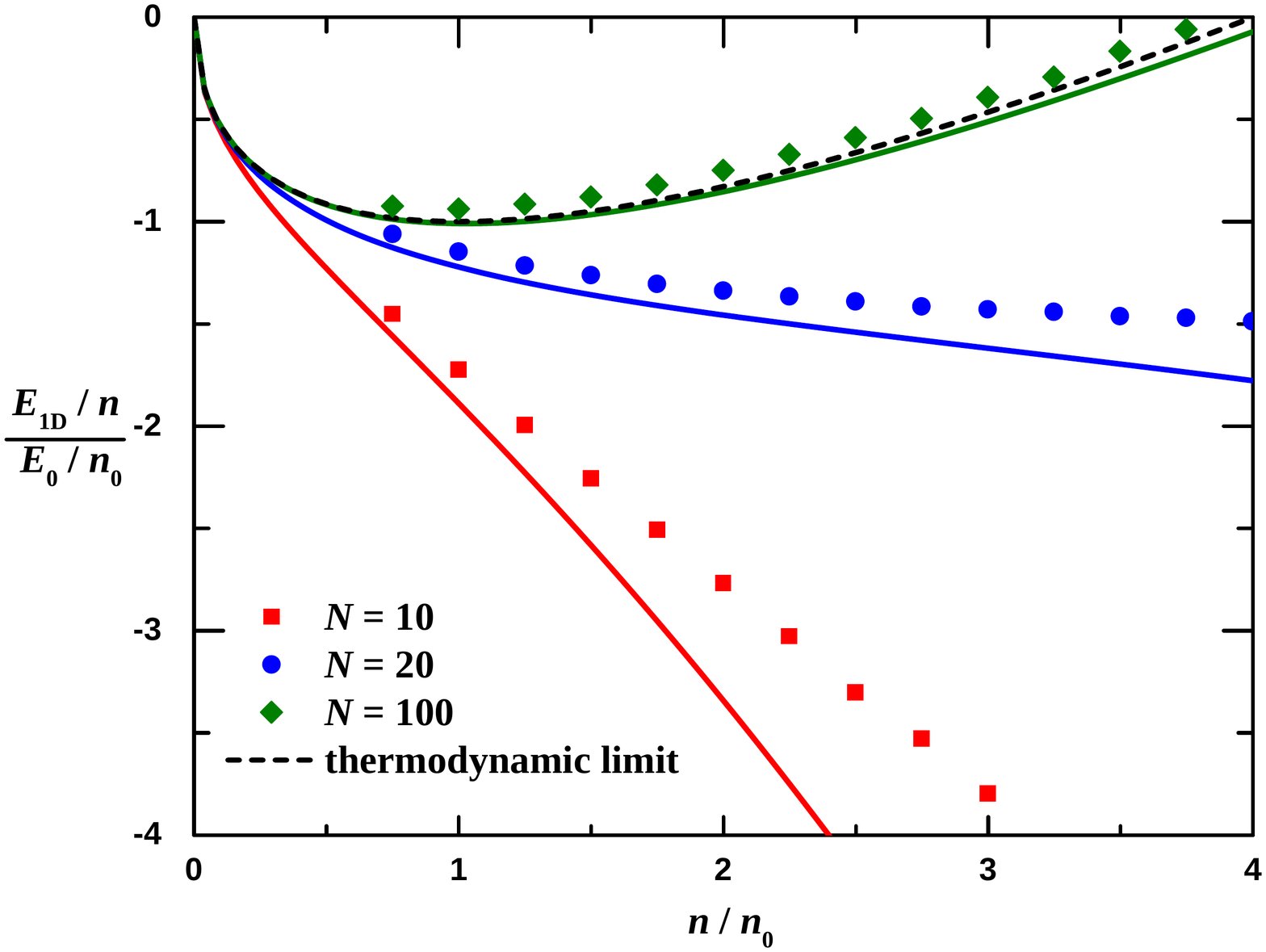}
\caption{The energy per particle $E_{\rm 1D}/2n$ versus density $n$ for the symmetric one-dimensional mixture with delta-function interactions and $\delta g/g = 0.05$ in a box with periodic boundary conditions with particle numbers in each component equal to $N = 10$ (red squares), 20 (blue circles), and 100 (green diamonds). Solid lines correspond to the Bogoliubov theory prediction with the leading-order finite-size correction, Eq.~(\ref{Eq:Bog1Dsymm:finitesize}), which can be rewritten as
$(E/n)/(E_0/n_0) = n/n_0 - 2\sqrt{n/n_0} - (2/9) (n/n_0)^{3/2} (g/N\delta g)^2$. Dashed curve is the Bogoliubov theory prediction in the thermodynamic limit.
}
\label{Fig:1Dfinitesize}
\end{figure}
\par\end{center}

\section{Finite-range effects\label{Sec:finite range}}

\begin{center}
\begin{figure}[ht]
\vskip 0 pt \includegraphics[clip,width=0.6\columnwidth]{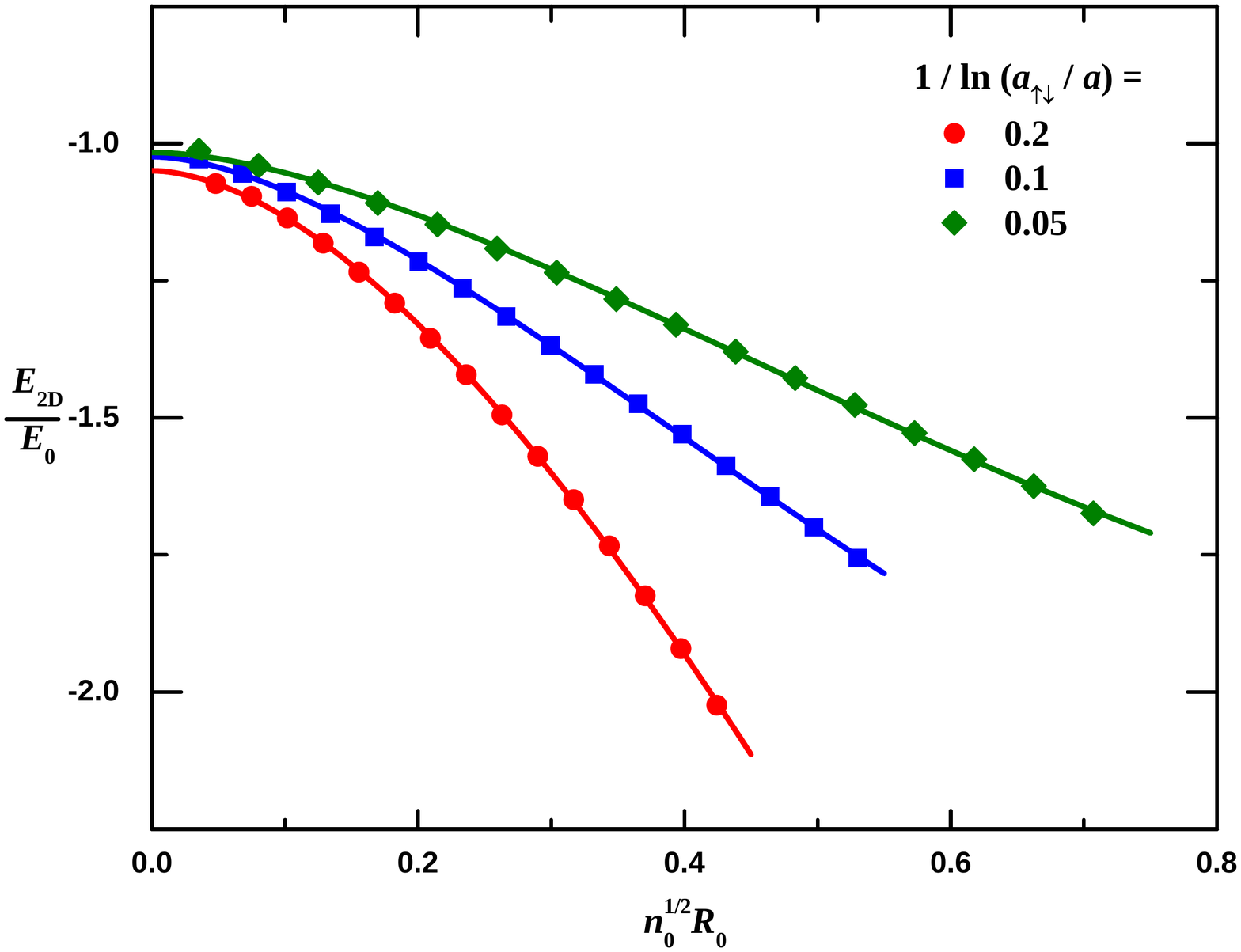}
\caption{
The energy per particle $E_{\rm 2D}/2n$ versus potential range $R_0$ for the symmetric two-dimensional mixture at $n=n_0$.
}
\label{Fig:finiterange}
\end{figure}
\par\end{center}

In order to study the dependence of the energy on the interaction range we perform calculations of the symmetric two-dimensional mixture with soft-disk repulsive intraspecies and square-well attractive interspecies interaction potentials of variable range $R_0$ but for fixed scattering lengths. Figure~\ref{Fig:finiterange} shows the energy per particle at the density $n=n_0$ as a function of $R_0$ for $1/\ln (a_{\1\2}/a) = 0.2$ (red circles), 0.1 (blue squares), and 0.05 (green diamonds).
Solid lines in Fig.~\ref{Fig:finiterange} are obtained by fitting the data with the empirical expression $E(R_0) = E [1 + A nR_0^2 \ln(B n^{1/2} R_0)]$, where $A$ and $B$ are fitting parameters. Note that the Bogoliubov zero-range theory prediction is recovered for $1/\ln (a_{\1\2}/a)\to 0$ even for fixed $n^{1/2}R_0$. Results presented in the main text are obtained for $nR_0^2 = 5\times 10^{-3}$, their difference from the zero-range asymptotic value is comparable to the symbol size. In one dimension both analytical and numerical approaches use contact $\delta$-function potential.

\end{widetext}
\end{document}